\documentclass[a4paper,11pt]{article}
\usepackage[english]{babel}
\usepackage{amssymb,amsmath,amsfonts,amsthm}

\usepackage{geometry}
\usepackage{graphicx}
\usepackage{xcolor}
\usepackage{tabularx}
\usepackage{float}
\usepackage{rotating}
\usepackage{multirow}
\usepackage{ifthen}
\usepackage{fancyhdr}
\usepackage[round,comma]{natbib}

\usepackage{setspace}

\let\OLDthebibliography\thebibliography
\renewcommand\thebibliography[1]{
	\OLDthebibliography{#1}
	\setlength{\parskip}{0pt}
	\setlength{\itemsep}{0pt plus 0.3ex}
}

\usepackage[hidelinks,backref=page,colorlinks=false,pdfborder={0 0 0},hyperfootnotes=false,breaklinks=true]{hyperref} 
\hypersetup{
pdfstartpage=1,
pdfstartview={XYZ null null 1},
pdfcenterwindow=true,
pdfborder={0 0 0},
colorlinks=false,
pdftitle={Human Life Is Unlimited - But Short},
pdfauthor={Holger Rootzen and Dmitrii Zholud},
pdfsubject={Applied Statistics, Extreme Value Theory},
pdfkeywords={Extreme human life lengths, no influence of lifestyle on survival at extreme age, no influence of genetic background on survival at extreme age, future record ages, supercentenarians, Jeanne Calment, limit for human life span; force of mortality; size-biased sampling; generalized Pareto distribution.},
pdfcreator={MiKTeX 2.9}}

\begin{document}
\hypersetup{
	pdfstartpage=1,
	pdfstartview={XYZ null null 1},
	pdfcenterwindow=true,
	pdfborder={0 0 0},
	colorlinks=false,
	pdftitle={Human Life Is Unlimited - But Short},
	pdfauthor={Holger Rootzen and Dmitrii Zholud},
	pdfsubject={Applied Statistics, Extreme Value Theory},
	pdfkeywords={Extreme human life lengths, no influence of lifestyle on survival at extreme age, no influence of genetic background on survival at extreme age, future record ages, supercentenarians, Jeanne Calment, limit for human life span; force of mortality; size-biased sampling; generalized Pareto distribution.},
	pdfcreator={MiKTeX 2.9}}	
\setcounter{page}{1}
\newtheorem{theorem}{Theorem}[section]
\newtheorem{lemma}{Lemma}[section]
\newtheorem{corollary}{Corollary}
\renewcommand{\abstractname}{\textsc{Abstract}}
\renewcommand{\refname}{\textbf{References}}
\renewcommand{\thecorollary}{\arabic{section}.\arabic{lemma}.\arabic{corollary}}
\renewcommand{\proofname}{\textbf{Proof:}}
\renewcommand{\qedsymbol}{$\Box$}
\renewcommand{\headrulewidth}{0pt}
\renewcommand{\footrulewidth}{0pt}
\long\def\symbolfootnote[#1]#2{\begingroup%
\def\thefootnote{\fnsymbol{footnote}}\footnote[#1]{#2}\endgroup}
\newcommand{\authorname}{Dmitrii Zholud}
\colorlet{HiddenColorCR}{white}
\newcommand{\hiddentext}[1]{\textcolor{HiddenColor}{#1}}
\newcommand{\hiddencopyright}[1]{\hypersetup{urlcolor=HiddenColorCR}\textcolor{HiddenColorCR}{#1}\hypersetup{urlcolor=urllinkcolor}}
\newcommand{\setwatermarkfoot}{%
\fancyfoot[L]{
          \ifthenelse{\isodd{\value{page}}}
          {}
          {\parbox{4.3cm}{\centering\tiny
            \hiddencopyright{Copyright \raisebox{0.1mm}{$\copyright$} \the\year \ by  \authorname\\
 	   \href{http://www.zholud.com}{www.zholud.com}}
          }}
          }%
\fancyfoot[R]{
	 \ifthenelse{\isodd{\value{page}}}
          {\parbox{4.3cm}{\centering\tiny
            \hiddencopyright{Copyright \raisebox{0.1mm}{$\copyright$} \the\year \ by  \authorname\\
 	   \href{http://www.zholud.com}{www.zholud.com}}
          }}
          {}
          }%
}
\newcommand{\articlenote}
{
\vspace{\fill}
\
\\
\fbox{
\parbox{16.5cm}{\tiny
This is an author-created version of the article which can be used for educational or research purposes, and to create derivative works,
subject to appropriate attribution and non-commercial exploitation.
The \textcolor{blue}{\href{http://www.zholud.com/articles/Human-life-is-unlimited-but-short.pdf}{\bf latest version}} is available at \href{http://www.zholud.com}{www.zholud.com}.
}
}
}

\title{{\href{http://www.zholud.com/articles/Human-life-is-unlimited-but-short.pdf}{Human life is unlimited -- but short}}}
\author{\\ Holger Rootz\'{e}n\thanks{\noindent
    \textit{Department of Mathematical Statistics}\newline
    \textit{Chalmers University of Technology and University of G\"{o}teborg, Sweden.}\newline
    \ \ \
    E-mail: \href{mailto:hrootzen@chalmers.se}{hrootzen@chalmers.se}} \ \ and \ Dmitrii Zholud\thanks{\noindent
    \textit{Department of Mathematical Statistics}\newline
    \textit{Chalmers University of Technology and University of G\"{o}teborg, Sweden.}\newline
    \ \ \
    E-mail: \href{mailto:dmitrii@chalmers.se}{dmitrii@chalmers.se}}
         }
\date{}
\maketitle
\fancyhead{}
\fancyfoot{}
\fancyfoot[C]{}
\fancyhead[L]{\small 
}
\thispagestyle{fancy}

\maketitle

\begin{abstract}
	\noindent
	Does the human lifespan have an impenetrable  biological upper limit which ultimately will stop further increase in life lengths? This question is important  for understanding  aging, and for  society, and has led to intense controversies.  Demographic data for  humans  has been interpreted as showing existence of a limit, or even as an indication of a decreasing limit, but also as evidence that  a limit does not exist. This paper studies what can be inferred from data about human mortality at extreme age. We show that in western countries and Japan and  after age 110 the probability of dying  is about 47\% per year.  Hence there is no finite upper limit to the human lifespan. Still, given the present stage of biotechnology,  it is unlikely that  during the next 25 years anyone will live longer than 128 years in these countries. Data, remarkably, shows  no  difference in mortality  after age 110 between sexes,  between ages, or between  different lifestyles or genetic backgrounds. These results, and the analysis methods developed in this paper, can help testing biological theories of ageing and aid confirmation of success of efforts to find a cure for ageing.
\end{abstract}

\section{Does the human lifespan have a finite limit?}

Inferences about mortality at extreme age crucially depend  on the quality of data, and on  analysis respecting the way data has been collected. Bad data or wrong analysis can, and has, led to misleading conclusions, and also good analyses become outdated when more informative data is collected. Conclusions about a limit for the human  lifespan are only meaningful if they are accompanied by a clear definition of the concept of a limit for the random variation of lifespans of individuals. They have led to intense controversies \citep{Couzin2011}.

A recent Nature Letter, \cite{dongetal2016},  claims that ``the maximum lifespan of humans is fixed and subject to natural constraints''. The claim rests on plots showing that the rate of improvement in survival  is slower at very high ages; the observation that annual average age at death of supercentenarians (humans who are  more than 110 years old) has not changed; and on analysis of the yearly highest  to the fifth highest reported age at death. However, a slower rate of improvement is still an improvement, and, if anything, it contradicts the existence of a  limit.  Further, record age will increase as more supercentenarians are observed, even if the distribution of their life lengths does not change.  Finally, the analysis of the yearly highest reported ages of death is misleading.  Thus the conclusion in \cite{dongetal2016} is unfounded and based on inappropriate use of statistics. Additional details are given in Section~\ref{sec:dong}.

A similar claim  that there is a ``biologic barrier limiting further life-span progression'' is made in \cite{Antero-Jacquemin}. The claim  is  based on analysis methods which do not distinguish between a finite and an infinite limit, which do not take truncation caused by sampling scheme into account (see discussion below), and   which sometimes exclude  the most extreme life lengths from analysis.  Neither of the papers \cite{dongetal2016} or \cite{Antero-Jacquemin}  define the concept of a limit for the human lifespan.

Earlier papers, \citep{weon-et-a.:2009,Aarssen-dehaan1994,wilmoth:2000}, argue both for and against a limit for the human lifespan. The arguments are based on analysis of  data from the Netherlands or Sweden with maximum recorded ages 112 years or less.  Extrapolation to conclusions about the existence, or not,  of a limit of 122.45 or more years, the age of the  longest living documented human, Jeanne Calment,  are long, and thus uncertain.
\pagestyle{fancy}%
\fancyhead{}%
\setlength{\headheight}{12pt}
\fancyhead[C]{
	\ifthenelse{\isodd{\value{page}}}
	{\small Human life is unlimited -- but short}
	{\small Holger Rootz\'{e}n and Dmitrii Zholud}
}%
\fancyfoot{}%
\fancyfoot[C]{{\thepage}}%
\setwatermarkfoot%

In a careful analysis \cite{gampe2010} studies survival after age 110 using  non-parametric EM  estimates in a life-table setting. Her results  agree with the present paper and hence, specifically, provide important confirmation that the parametric models we use are appropriate. However our  parametric approach  makes a much more detailed analysis possible,  and, in particular, addresses the question of a limit for the human lifespan; make predictions about future record ages possible; and makes it possible to perform  tests and to obtain reasonable confidence intervals.  There is some  consensus that ``the notion of an intractable species-specific senescent death and species-specific maximum lifespan'' has been refuted \citep{vaupel2010}.

To summarize, the question if there is a limit to the human lifespan still has not received any convincing answer. In this paper we use the best available data, the International Database on Longevity \citep{IDL2016}  to give the currently best possible  answer to the question, and  to study how factors such as  gender, age,  lifestyle, or genetic background change - or do not change - human mortality at extreme age. We use statistical Extreme Value Theory which is the appropriate statistical framework for study of extreme ages. 

Unvalidated data on ages of supercentenarians  is known to be completely unreliable \citep{poulin2010}. The IDL database  contains validated life lengths  of 668 supercentenarians from 15 countries, with ``data collection performed in such a way that age-ascertainment bias is avoided''. This means that validation was performed with equal effort for younger and older supercentenarians. The  only other data on  validated lifespans of supercentenarians, the \cite{grg2016} list,  does not have a clearly specified plan for data collection, and is expected to be agebiased.

The sampling scheme in the IDL database is somewhat different for different countries, but typically IDL contains ages of persons who died in some time interval, say 1980-1999, at age exceeding 110 years. The time intervals are different for different countries. This sampling scheme leads to both left and right truncation of life lengths. The analysis in this paper throughout takes this truncation into account; for details see Section~\ref{sec:bias}. This is important: analysis which does not do this properly would lead to crucially different results.

Extreme value statistics  provides methods for analysis of the most extreme parts of data: extreme floods, extreme winds, extreme financial events -- or extreme life lengths, here excess life length after age 110. The generalized Pareto (GP) distribution plays the same role for analysis of extreme excesses as the Gaussian distribution does in other parts of statistics \citep{coles2001,beirlant2004}.

We summarize the main results of the analysis in Section~\ref{sec.unbounded}. Details on data and statistical analysis are given in Section~\ref{sec:methods}, and comments on the \cite{dongetal2016} paper in Section~\ref{sec:dong}. Section~\ref{sec:discussion} contains a concluding discussion.

\section{The human lifespan is unbounded, but short}
\label{sec.unbounded}

\begin{figure}[H]
	\centering
	\includegraphics[width=0.3\textwidth]{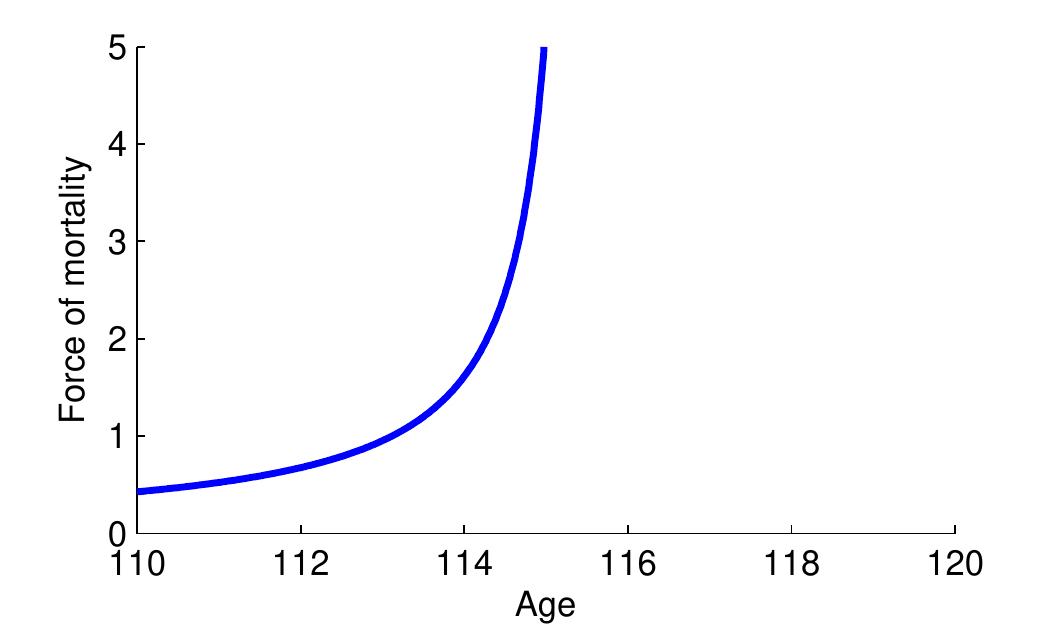}
	\includegraphics[width=0.3\textwidth]{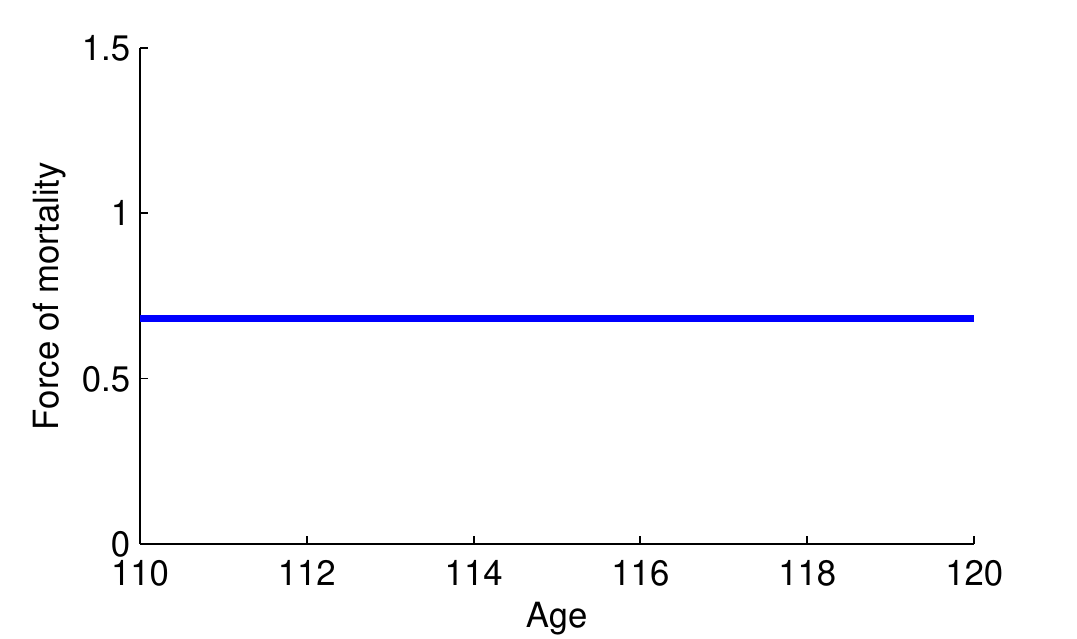}
	\includegraphics[width=0.3\textwidth]{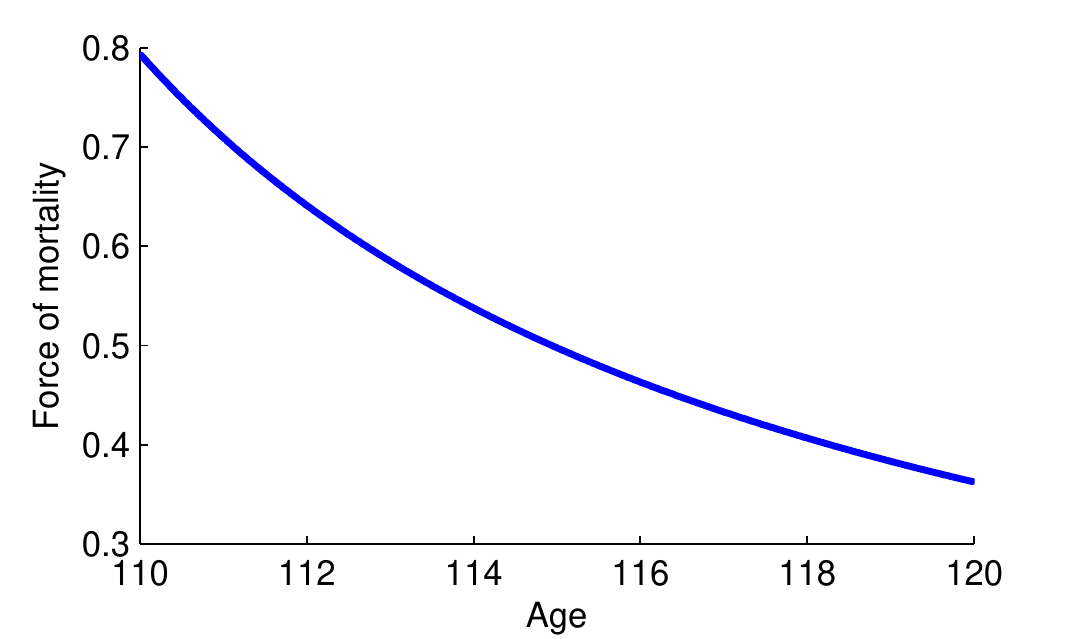}
	\caption{  {\em Left}:  GP distribution fitted to  IDL data for Japan, without taking truncation into account. Life  appears to have a finite limit equal to 116.0 years. {\em Middle}: Exponential distribution  fitted to all  IDL data with validation level A, with truncation taken into account. Life is unlimited but short. {\em Right}:   GP distribution fitted to  IDL data  for France, without truncation taken into account. Life appears to be unlimited. However, changing estimation for Japan and France so that truncation is taken into account leads to the conclusion that, in fact, also in these countries life is unlimited but short.}
	\label{fig:forceofmortality}
\end{figure}

Our aim is to estimate the ``force of mortality'' at extreme age.  The force of mortality, or hazard rate,  is the instantaneous rate of mortality   so that, e.g., (assuming the measurement unit is year) the probability of dying tomorrow is 1/365 times the force of mortality today. Formally, the force of mortality equals minus the derivative of the log survival function (here again with survival measured in years). We model excess life length (life length minus 110 years) of supercentenarians with a GP distribution.  The GP model includes three different cases with  distinct behaviors of the force of mortality. In the first case there is a finite age at which the force of mortality tends to infinity and beyond which survival is not possible -- in this case lifespans have a finite limit; in the second case the force of mortality is constant -- life is unlimited but short (more precisely, survival after age 110 is short); and in the third case the force of mortality decreases with age -- life is unlimited. In the second case excess life length has an exponential distribution. The three cases are illustrated in Fig.~\ref{fig:forceofmortality}. Additionally, Fig.~\ref{fig:forceofmortality}  shows the importance of taking truncation caused by the sampling scheme into account. 

Likelihood ratio tests using the assumption of a GP distribution, or assuming an exponential distribution,  show that although around 10 times as many women as men live to age 110 years, differences in  mortality, if any,  between women and men after age 110 are too small to be detectable from the IDL data.
Likelihood ratio tests also did not detect any difference in survival between south and north Europe, between western Europe and north America, or between Japan and the other countries. Wald tests did not identify any differences in mortality above age 110 for earlier and later parts of the data, and likelihood ratio tests did not reject the hypothesis that survival after age 110 has an exponential distribution. The resulting model,  that mortality after 110 years of age is constant, fits the data well (Tables ~\ref{table:women-men}-\ref{table:test-exp}, Fig.~\ref{fig:dataandqq}).

Thus, our conclusion is that in western Europe, north America, and Japan excess human life length after age 110 follows an exponential distribution -- human life is unlimited, but short -- and that survival is the same for women and men, during earlier periods and later ones, and in countries with very different lifestyles and genetic compositions. The mean of the exponential distribution was estimated to be 1.34 with 95\% confidence interval (1.22, 1.46).   Survival  after 110 in these countries can hence be described as follows: each year a coin is tossed, and if heads comes up it means a person will live one more year; more precisely, the estimated probability to survive one more year is 47\%, with  95\% confidence interval (0.44, 0.50).

\begin{figure}[H]
	\centering
	\includegraphics[width=0.49\textwidth]{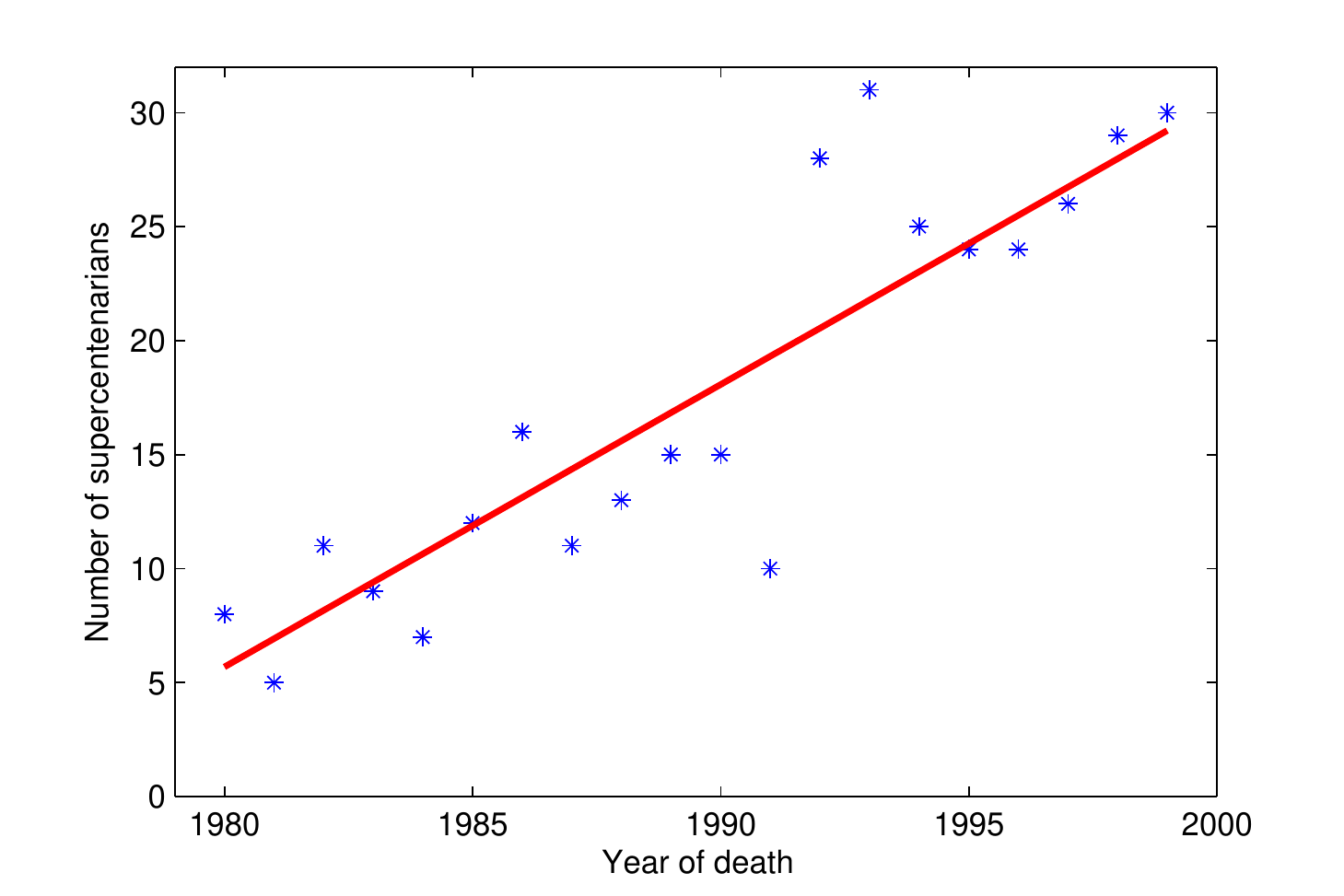}
	\includegraphics[width=0.49\textwidth]{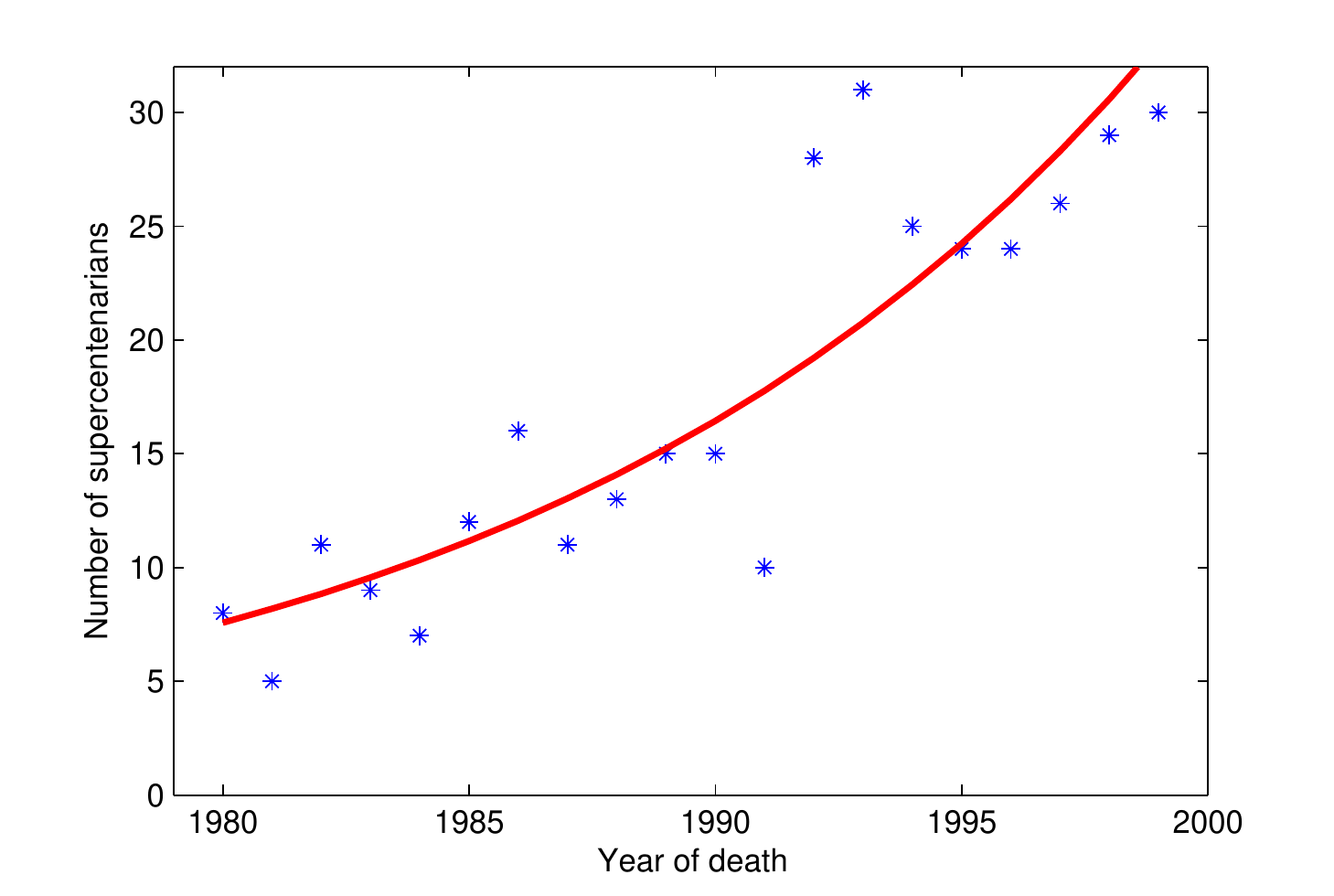}
	\caption{Yearly number of deaths of validated supercentenarians from  England and Wales, Italy, and USA. Line is computed using  Poisson regression with {\em Left:} identity link {\em Right:} log link.}
	\label{fig:regression}
\end{figure}

\begin{figure}[H]
	\centering
	\includegraphics[width=0.48\textwidth]{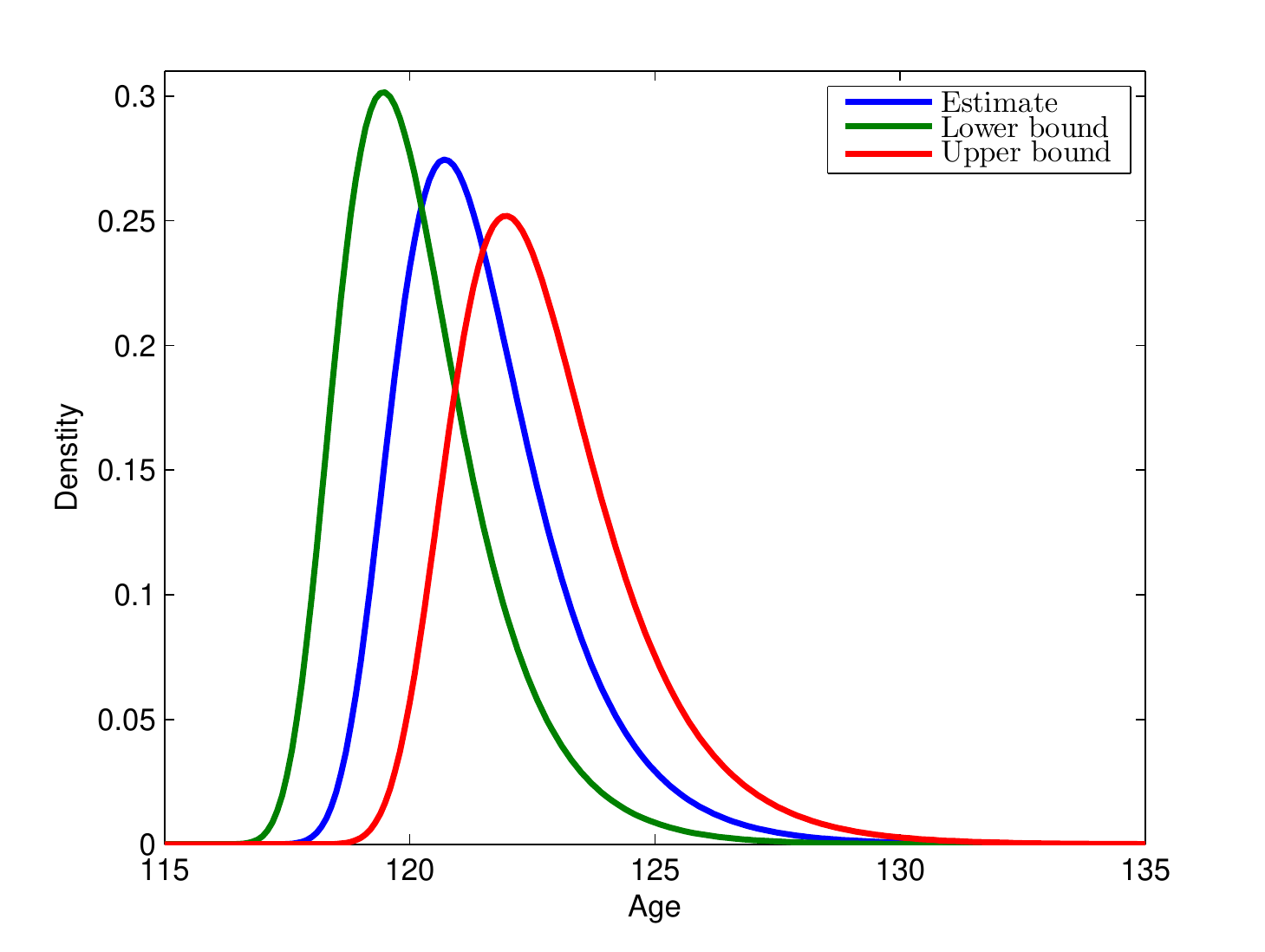}
	\includegraphics[width=0.48\textwidth]{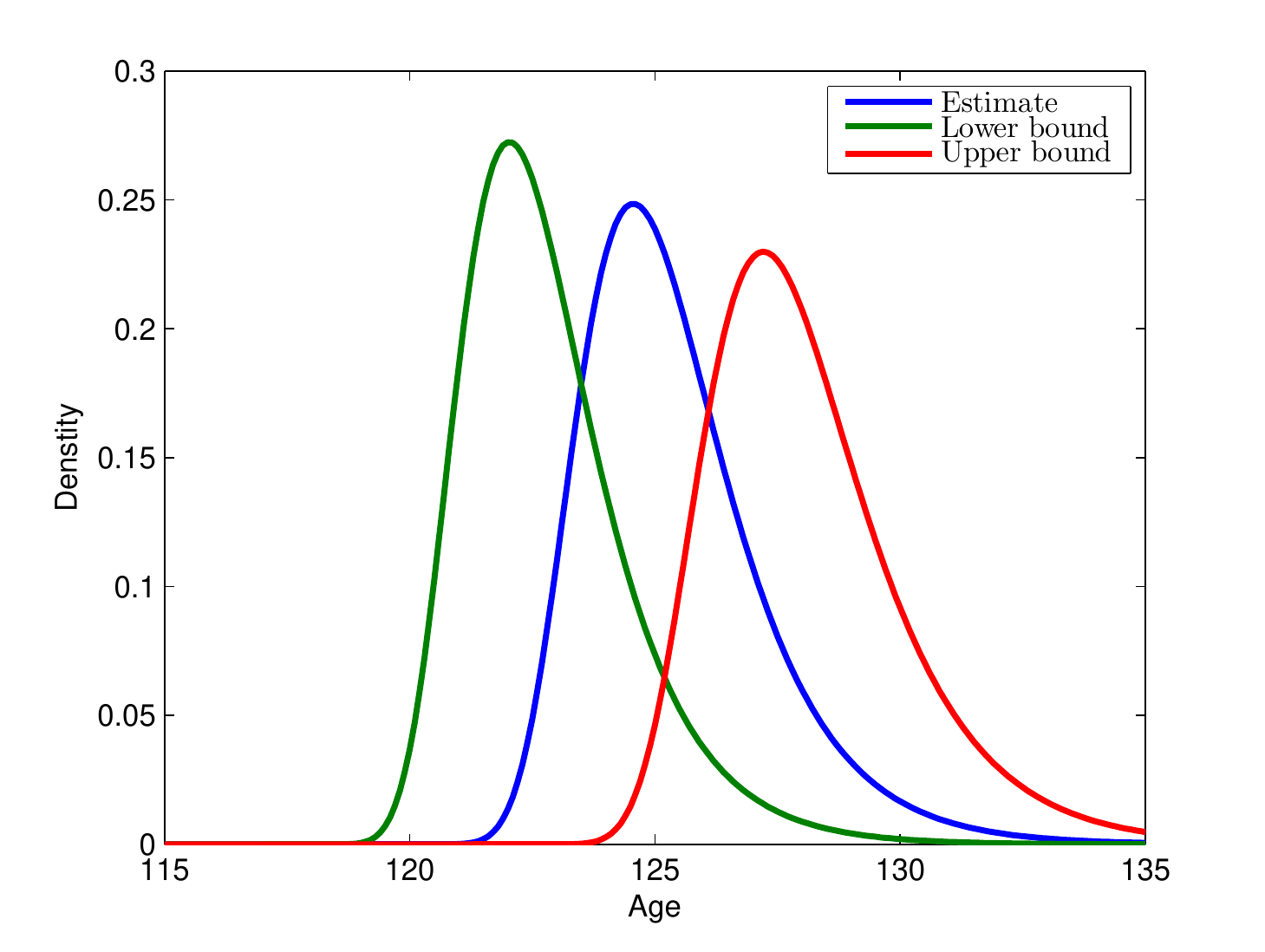}
	\caption{ Probability density for record life length during 2018-2042 moves left if parameter  $\sigma$ of exponential distribution, or the number of deaths of supercentenarians ($n$), are decreased, and to the right if they are increased.  Plots show estimated density of record life length (blue) together with lower (green) and  upper (red) bounding densities obtained by setting  $\sigma$ and $n$ equal to the lower or upper bounds in their confidence intervals. Parameter $n$ is estimated using  Poisson regression with {\em Left}: identity link {\em Right}: log link.}
	\label{fig:record}
\end{figure}

The number of persons who live longer than 110 years is rapidly increasing (Fig.~\ref{fig:regression}). Still, it is a very extreme event that a woman  will live  110 years: the probability is only about 2 out of 100,000. For a man this probability is ten times smaller. Given an unchanged mortality after age 110 and a continued increase in the number of supercentenarians, it is likely that the record age which will be documented in western Europe, north America or Japan during the coming 25 years will exceed 119 years but will be shorter than 128 years (Fig.~\ref{fig:record},  Section~\ref{sec:statistics}). If the record falls outside this interval, it would be an indication that survival has changed.


Jeanne Calment lived 3.18 years longer than the second longest-living person in the IDL data, which seems extreme, and could indicate that she is an ``outlier''. Her life length is the longest of the 566 supercentenarians with validation level A, and the  probability that the largest of 566 exponential variables with mean 1.34 is larger that 12.45 (her excess life length, 122.45-110) is 5\%. Thus there is some evidence that her life length is close to being an outlier -- or that, alternatively, mortality decreases at extremely high age, say after age 118.

\section{Materials and Methods}
\label{sec:methods}
\subsection{International database on longevity}
The \cite{IDL2016} database contains validated supercentenarian life lengths from 15 countries. The inclusion criterion typically is that the death has occurred in a specified time interval, however with different time intervals for different countries, and with somewhat more complicated criteria for USA and Japan. There are two validation levels, A and B, with level A being the most thorough validation.

In the analysis of the IDL data we only used the 566 life lengths with validation level A, and  countries were grouped geographically, and so that groups contain 64 or more supercentenarians, see Table \ref{table:groups}. The grouping was mandated by the fact that data from several countries included too few supercentenarians to make it meaningful to analyse these countries separately.
\begin{table}[H]
	\centering
	\caption{Data sets used for analysis. IDL data with validation level A. Persons who died after December 31, 1999, in USA, or died in 1996 or after August 31, 2003, in Japan, are excluded.}
	\label{table:groups}
	\vspace*{3mm}
	\begin{tabular}{l  l c c}
		\hline
		ID & consists of &  women & men
		\\
		\hline
		North Europe & Denmark, Germany, England and Wales & 79 & 5 \\
		South Europe & Italy, France, Spain & 97 &17 \\
		Europe & North and south Europe & 176 & 22\\
		North America & USA, Quebec & 272    & 28 \\
		Japan & Japan & 55 & 9 \\
		Europe\&America &   Australia, Europe, North America, & 452 & 50 \\
		World & Europe\&America, Japan & 507 & 59\\
	\end{tabular}
	
\end{table}
We excluded persons who died after December 31, 1999 in the USA,  and persons who died in 1996 or after August 31, 2003 in Japan; see Section~\ref{sec:bias} below.  In the sequel we for brevity just write ``Europe\&America'' for the data set which contains countries from western Europe,  north America, and Australia, and ``World'' for the entire data set. To investigate the possibility of time trends in mortality of supercentenarians  the data sets were split into two parts according to year of death. Except for World, the division was made so that the second part included more persons than the first one, but with the difference between the number of persons in the parts as small as possible.
\begin{figure}[H]
	\centering
	\includegraphics[width=0.49\textwidth]{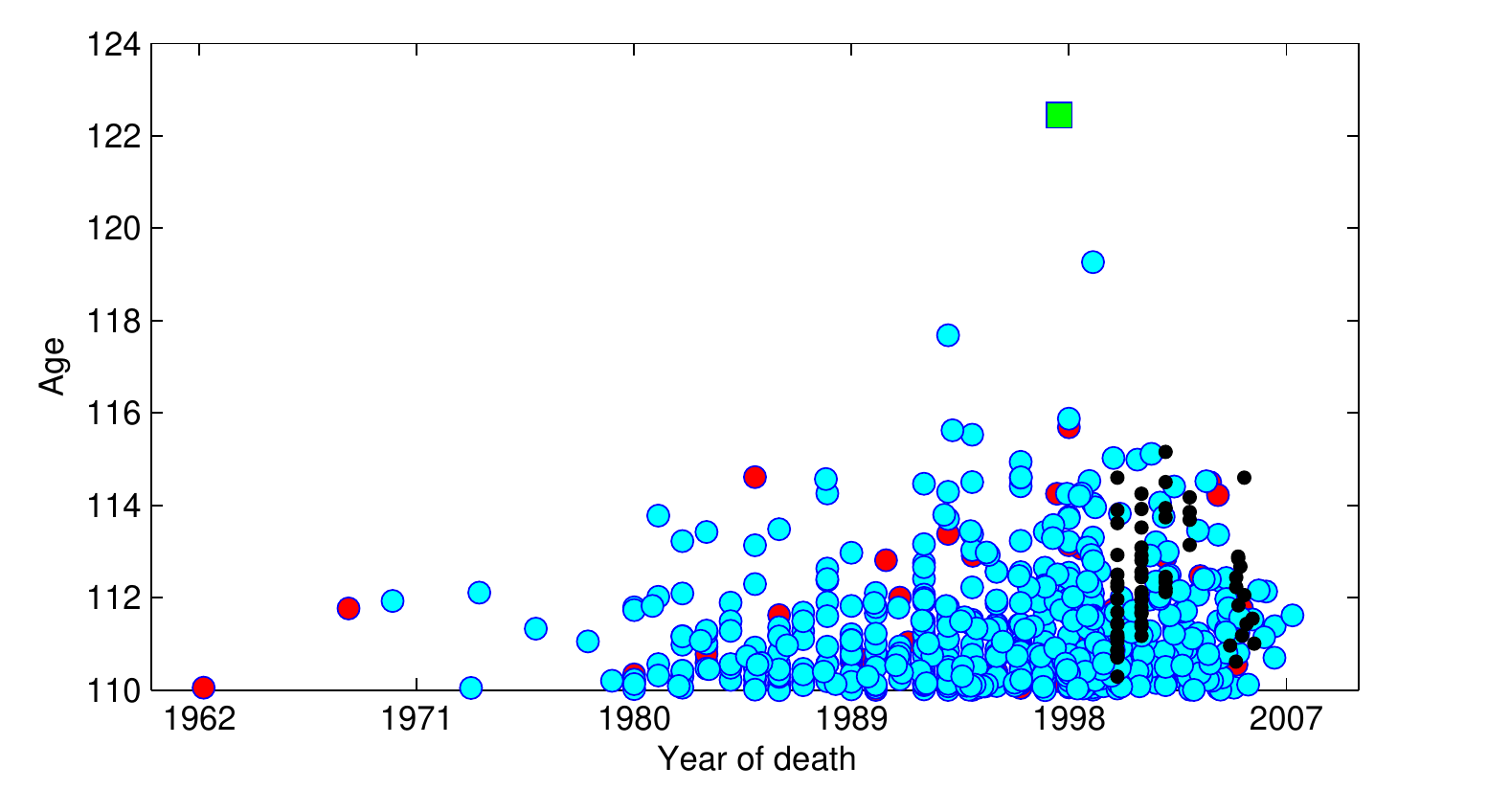}
	\includegraphics[width=0.49\textwidth]{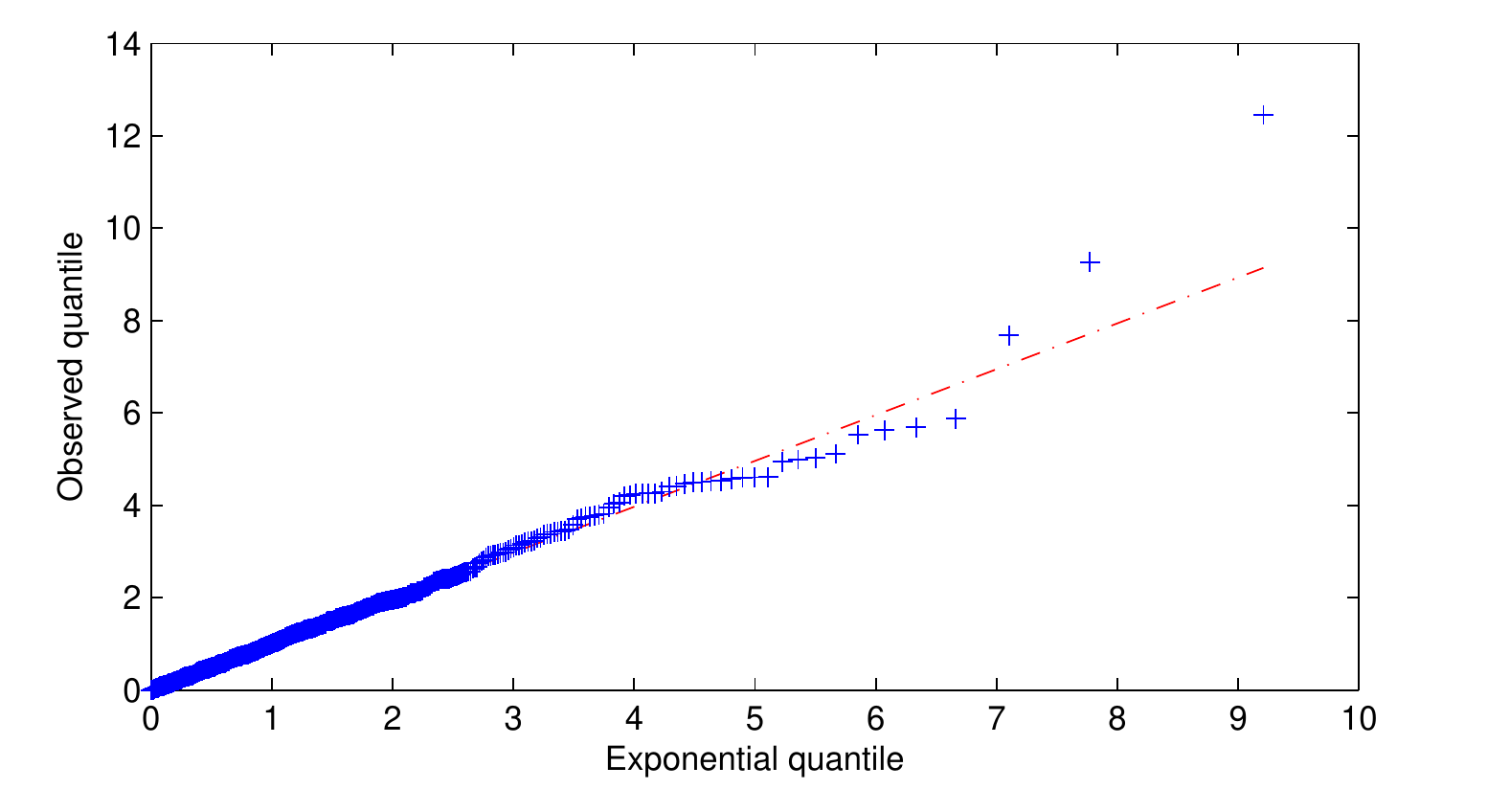}
	\caption{IDL data with validation level A, from ``"World''. {\em Left:} Life lengths against year of death: blue - women, red - men, black - persons who died after December 31, 1999 in USA or in 1996 or after August 31, 2003 in Japan and which are not included in analyses. Green square is Jeanne Calment. {\em Right:} Exponential qq-plot. }
	\label{fig:dataandqq}
\end{figure}
Including the $32$ persons in countries validated at  level B  (Belgium, Finland, Norway, Sweden, Switzerland) in the analysis did not change conclusions. We also did the analyses reported below for life lengths in excess of higher age thresholds: 110.5, 111, 111.5, and 112 years. This did not change any of the conclusions, although estimates of course were somewhat different, and confidence intervals were wider for higher thresholds. The estimates and tests obtained after including  validation level B data and/or using higher thresholds are given by the MATLAB toolbox LATool which we have developed and which is available as Supplementary Material. Similarly, removing the largest age, Jeanne Calment's age, from the data did not change conclusions.

\subsection{Statistical analysis}
\label{sec:statistics}
The GP distribution has cumulative distribution function $G(x)=1-(1+\gamma x / \sigma)_+^{-1/\gamma}$ and force of mortality function $(1+\gamma x / \sigma)_+^{-1}/\sigma$, where $+$ signifies that the expression in parentheses should be replaced by zero if negative. Here $\gamma$ is a shape parameter, termed the extreme value index, and $\sigma$ is a scale parameter. The three different cases, limited life length, unlimited but short life length, and unlimited life length, correspond to $\gamma<0, \gamma=0$ and $\gamma>0$, respectively.

The statistical analyses were  done using LATool, the  MATLAB toolbox which is available as Supplementary Material. The toolbox uses maximum likelihood estimation which takes the truncation caused by the sampling scheme into account, see Section~\ref{sec:bias}. Confidence intervals were computed using asymptotic normality.  The two sets of p-values in Tables \ref{table:women-men} and \ref{table:countries} correspond to the two possible testing strategies: first test using the GP distribution, and then in the end test for $\gamma=0$, i.e. for an exponential distribution, or conversely start with testing for an exponential distribution and then do the other tests assuming an exponential distribution. Both strategies lead to the same conclusions.

The probability to survive one more year is $e^{-1/\sigma}$. Inserting the estimate and confidence limits for $\sigma$ into this formula gives the estimate and confidence interval for the yearly survival probability.

The maximum, or record,  of $n$ exponential random variables with parameter $\sigma$ has cumulative distribution function $(1- e^{-x/\sigma})^n$ and probability density function $ne^{-x/\sigma}(1- e^{-x/\sigma})^{n-1}/\sigma$. Hence, to predict the size of the record age in some time period both the value of $\sigma$ (estimated above) and the value of $n$, the number of supercentenarians who will die in the period, is needed.

To estimate the trend in the number of supercentenarians using data from different countries, the data must cover the same time period. Such data is available for for supercentenarians who died in the time period 1980-1999, for the countries Italy, England and Wales, and USA. Using  Poisson regression with linear link function, the expected number of supercentenarians   who will die in these three countries during  the next 25 years, i.e. in 2018-2042, was estimated to be 1,690 with a 95\% confidence interval (1,326, 2,054).

The ratio  of the number of   persons aged 100--104 in  north America, western Europe and Japan\footnote{Austria,  Canada, Denmark, Belgium, France, Finland, Iceland, Ireland, Italy, Japan, the Netherlands, Norway,  Portugal,  Spain, Sweden, UK, USA.} in the year 2000 in the Human Mortality Database \citep{humanmort2016} to the same number for Italy,   England and Wales, and  USA is 1.76.  We used this ratio, 1.76, as a proxy for the relative number of deaths of supercentenarians in these countries. Multiplying the estimate and confidence interval with 1.76 gave the estimate $n=2,974$, with confidence interval (2,334, 3,615), for the expected number of supercentenarians who will die in the period  2018-2042 in north America, western Europe and Japan. Assuming independence of this interval and the 95\% confidence interval for $\sigma$ gives a  joint confidence level 90\%.

The same computations using a log link (i.e. assuming an exponential growth of the number of supercentenarians) in the Poisson regression gave the estimate $n=18,717$ with confidence interval (7,429, 46,962) obtained by simulating parameter values from the limiting normal distribution. It may be noted that these predictions only depend on changes in survival up to age 110 and are not affected by future fluctuations in birth rates: all supercentenarians who will die in the interval 1918-1942 are already born.

Inserting $x= $ age$-110$ and the estimates $\sigma =1.34, n=2,974$   into the formula for the probability density function for the record  gives the blue graph in Fig.~\ref{fig:record}, left panel.
Instead using the lower confidence interval bounds $\sigma=1.22, n=2,334$ gives a lower bound  probability density function: this is the green graph in Fig.~\ref{fig:record}, left. The upper bound, and the graphs in Fig.~\ref{fig:record}, right, are computed similarly.

Using the formula for the cumulative distribution function of  record age and the linear estimate of $n$  gives  the estimate 3\%  for the probability that the record age is lower than 119. Instead inserting  the lower bounds of the confidence intervals into the formula gives the upper bound estimate  23\% of the probability. Similar computations, using the exponential growth estimate of $n$ gives the estimate 3\% for the probability that the record age exceeds 128 years, and the upper bound 19\% for this probability. This shows that it is likely that  the record age will fall in the interval 119-128 years. These results are obtained assuming that  present mortality and growth of number of supercentenarians remain unchanged.

There are only 10 persons in the World data set who lived longer than 115 years, and hence  little data to confirm an exponential distribution beyond this age. Still these 10 data points are in agreement with  an exponential distribution (Fig.~\ref{fig:dataandqq}). In view of predictions of very rapid growth of the number of  centenarians \citep{thatcher2010}, the results based on the exponential regression for $n$ may capture the most likely development.

In the computations above we for simplicity of exposition used the expected rather than actual, random, number of supercentenarians.  However, assuming a Poisson distribution   with mean $n$ for the number $N$ of supercentenarians who die in 2018-2042, the distribution of the record age is
\begin{eqnarray*}
	P(M_N \leq x) &=& E(P(M_N \leq x | N)) = E((1-e^{-x/\sigma})^N)  \\
	&=& e^{n\{(1-e^{-x/\sigma})-1\}} =  e^{-ne^{-x/\sigma}}.
\end{eqnarray*}
It can be seen by numerical computation that for the parameter values used above the difference between this expression and $(1-e^{-x/\sigma})^n$ is less than one unit in the second decimal. The same holds for the densities obtained by differentiation of the two formulas.

\subsection{The IDL sampling scheme: left and right truncation}
\label{sec:bias}
\begin{figure}[H]
	\centering
	\includegraphics[width=0.49\textwidth]{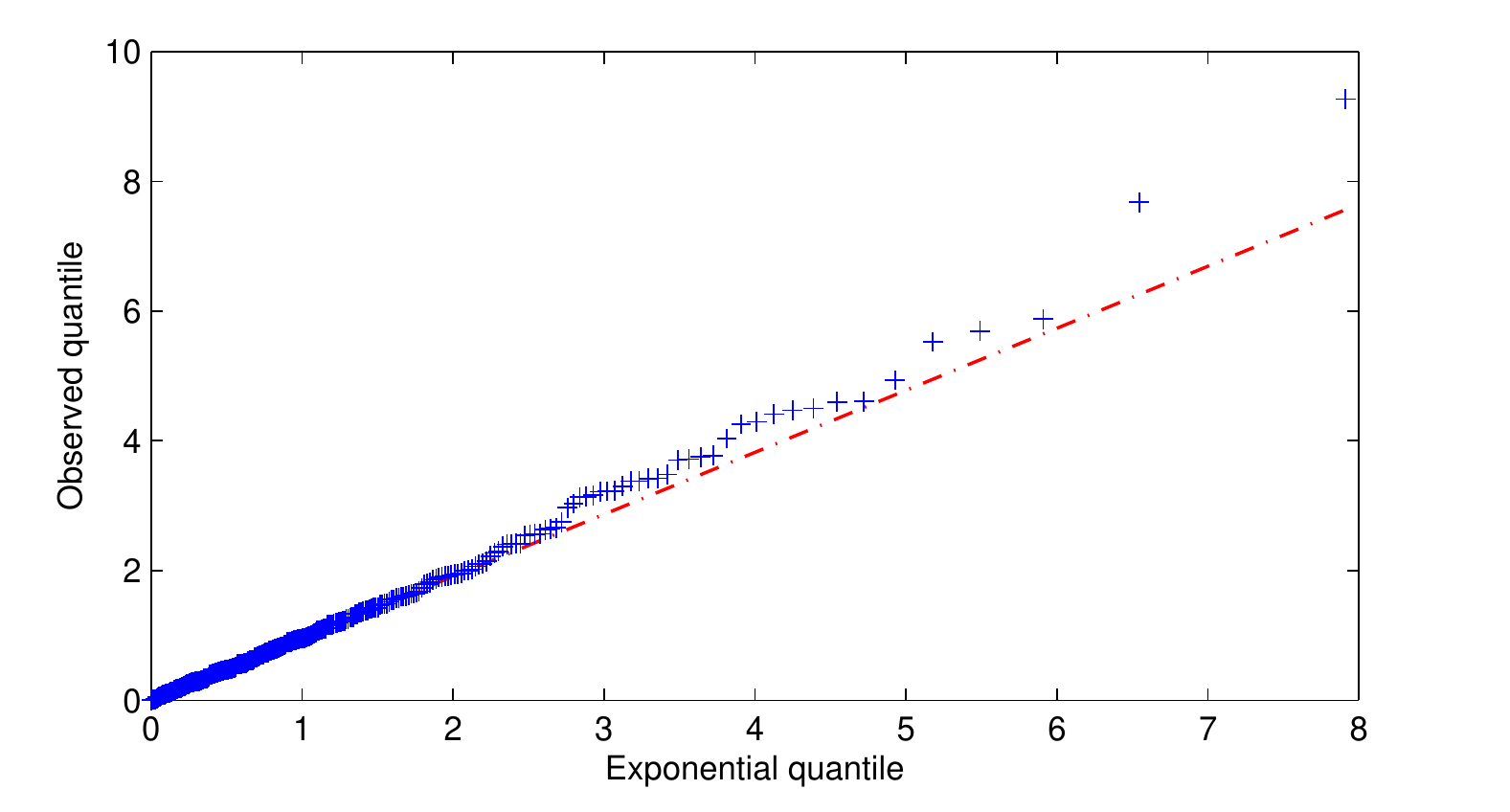}
	\includegraphics[width=0.49\textwidth]{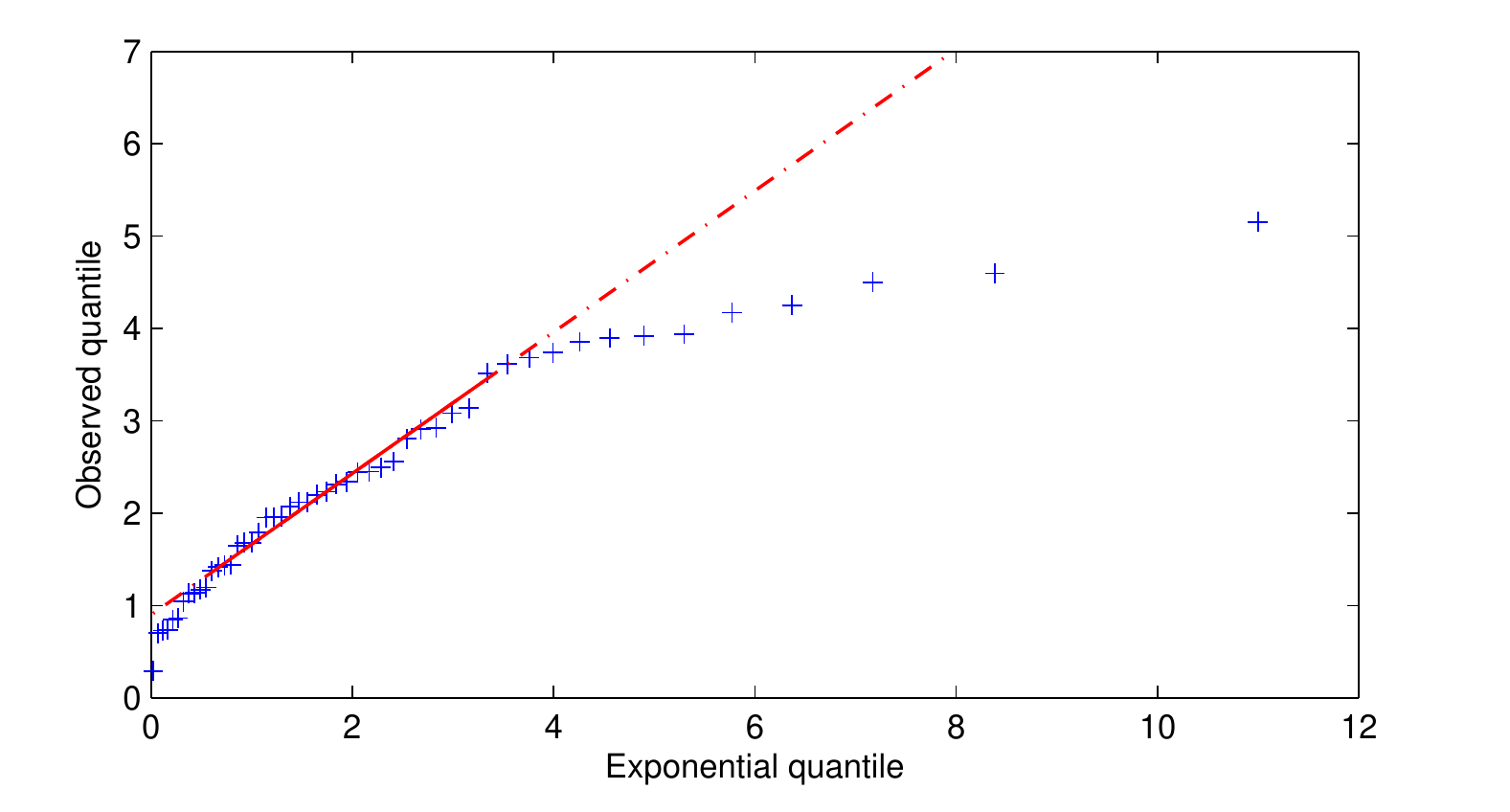}
	\caption{llustration of effect of size-biased sampling: {\em Left}: Exponential qq-plot for IDL life lengths  of supercentenarians who died 1980-1999 in  USA. {\em Right}: Exponential qq-plot for IDL life lengths  of supercentenarians who died 2000-2003 in  USA.}
	\label{fig:qq-bias}
\end{figure}

The two exponential qq-plots in  Figure~\ref{fig:qq-bias} are quite different, and illustrates that the sampling scheme has to be taken into account in the analysis. Specifically, the explanation of the difference between the plots is not that US mortality changed at the end of 1999, but instead that the sampling schemes for the 1980-1999 data and the 2000-2003 data were different. The 1980-1999 data consists of ages of supercentenarians who died in this interval, while the 2000-2003 data consists of supercentenarians who were alive at the end of 1999 but died before the middle of 2003. The effects of the 1980-1999 sampling scheme are discussed below. The 2000-2003 sampling scheme  leads to two kinds of biases:  persons who have longer lives are more likely  to be alive at any given point of time, say December 31, 1999, but on the other hand,  persons with very long life are not included in the data set because they have not died before the middle of 2003. These effects are different from those caused by the 1880-1999 sampling scheme, see below. Simulations confirmed that these sampling schemes can produce qq-plots like those in Fig. \ref{fig:qq-bias}. A further illustration of the importance of taking the sampling scheme for the IDL data into account is given by Figure~\ref{fig:forceofmortality}.

In principle, the bias caused by the different sampling scheme used for the USA 2000-2003 data could be taken into account, and also this part of the data could be used. However, the information given about the sampling plan for IDL supercentenarians who died in the USA after 1999 was not sufficient to make it possible for us to do this. Hence, supercentenarians who died in the USA after December 31, 1999, have been excluded from the analysis.  For similar reasons,  supercentenarians who died in Japan in 1996,  or after August 31, 2003, have also been excluded from the analysis.

Additionally, for reasons of confidentiality, for US supercentenarians only the year of death, and not the exact death date, is given. In the analysis we assumed that all US supercentenarian deaths occurred on July 2. However, conclusions didn't change if we instead set the death dates to January 1, or to December 31.

If the sampling scheme is that a cohort, all persons who are born in some time interval, is followed until all persons in it are dead, then there is no bias in the observed ages. Similarly, if a population is stable and there are no trends in the probability of becoming a supercentenarian, then recorded ages of supercentenarians who died in some time interval, say 1980-1999,  are not subject to  bias. However, the IDL database does not follow cohorts until they are extinct, and the probability of living until age 110 is increasing with time. This increase, if not taken into account, would lead to bias in the estimation of life length from the IDL data:  small excess life lengths are less likely to be included in the beginning of the interval but more likely to be included in the end of the interval -- and there are more deaths in the end. In the estimates and test performed here, we have taken this into account by letting the likelihood contribution from a person be conditional on the time of achieving age 110 and of death date being contained in the observation interval, as follows.

Suppose that a sample consists of all supercentenarian deaths in some country during the time interval $(b, e)$. Let $\{t_i\}$ be the times of achieving age 110 and $\{x_i\}$ be the corresponding excess ages at death. Further suppose that excess life lengths have cumulative distribution function $F$ and probability density function $f$, where in our analysis $F$ is either a GP distribution function or an exponential distribution function. Then supercentenarian $i$ is included in the sample if one of the following two conditions are satisfied: if a) $t_i \leq b$ and $b\leq t_i +x_i < e$ or if b) $t_i > b$ and $ t_i +x_i < e$.

In the case a) the excess life length of supercentenarian $i$ is truncated so that it only is included in the sample if  $b - t_i \leq x_i < e - t_i$. Hence the likelihood contribution from the excess life length of a supercentenarian in the sample whose 110 year birthday was before the beginning of the sampling time interval, i.e. with $t_i \leq b$, is
$$
\frac{f(x_i)}{F(e - t_i) - F(b-t_i)},
$$
In case b) where the 110 year birthday occurred after the beginning of the sampling interval, $t_i > b$, the excess life length is truncated to so that $x_i \leq e_i -t_i$, and  the likelihood contribution from supercentenarian $i$ is
$$
\frac{f(x_i)}{F(e - t_i)}.
$$
The full likelihood is then obtained by taking the product of the likelihood contributions for all supercentenarians and countries included in an analysis.

\begin{table}[H]
	\centering
	\caption{p-values for likelihood ratio tests of  no difference in mortality between women and men after age 110. Middle column assumes GP distribution, right column  exponential distribution.}
	\label{table:women-men}
	\vspace*{3mm}
	\begin{tabular}{l  c c}
		\hline
		ID  & GP & exp
		\\
		\hline
		North Europe & 0.94 & 0.81 \\
		South Europe & 0.14& 0.40 \\
		Europe & 0.57 & 0.63\\
		North America & 0.72     & 0.56\\
		Japan & 0.77 & 0.76\\
		Europe\&America &   0.42& 0.43 \\
		World & 0.43  &  0.41  \\
	\end{tabular}
\end{table}
\vskip -5mm
\begin{table}[H]
	\centering
	\caption{p-values for Wald tests of  no difference in mortality between first and last half of data. ``mean'' is the estimated parameter of the exponential distribution, not the mean of the observed life lengths; tests assume an exponential distribution.}
	\vspace*{3mm}
	\begin{tabular}{l c c c c c ccc}
		\hline
		ID & death date & \# & mean &death date & \# &mean& p-value
		\\
		\hline
		North Europe & 1970-1997& 41 & 1.78 &1998-2006 & 43  & 1.45 & 0.49    \\
		South Europe & 1973-1997 & 50 & 1.61 &1998-2007  & 64 & 1.19 &0.26  \\
		Europe & 1968-1997& 91 & 1.70 & 1998-2007 & 107  & 1.29 & 0.18 \\
		North America &1962-1992& 141 & 1.17  &1993-2002 & 159 & 1.14  & 0.73  \\
		Japan & 1996-2000  & 28  & 2.71 & 2001-2005 & 36& 1.28 & 0.30\\
		Europe\&America \hspace*{-3mm}& 1962-1994 & 242 & 1.34  & 1995-2008& 251 & 1.21 & 0.33 \\
		World & 1962-1996 & 309 & 1.30  & 1997-2008& 260 & 1.20 & 0.33  \\
	\end{tabular}
	\label{table:early-late}
\end{table}
\vskip -5mm
\begin{table}[H]
	\centering
	\caption{p-values for likelihood ratio tests of  no difference in mortality between groups of countries. Middle column assumes GP distribution, right column  exponential distribution.  A ``-'' means that the test for exponential distribution of data rejects at the 1\% level for  the GRG data set. GP distribution fits later part of GRG data badly.}
	\label{table:countries}
	\vspace*{3mm}
	\begin{tabular}{l c c}
		\hline
		ID & GP & exp\\
		\hline
		North Europe vs south Europe & 0.30 & 0.48 \\
		Europe vs north America & 0.24 & 0.23 \\
		Europe\&America vs Japan & 0.35 & 0.41 \\
		World vs GRG & 0.00     & -\\
	\end{tabular}
\end{table}
\vskip -5mm
\begin{table}[H]
	\centering
	\caption{Estimates  of shape parameter $\gamma$ of the generalized Pareto distribution,  with 95\% asymptotic confidence intervals in parenteses, and p-values for likelihood ratio test of the hypothesis that data follows an exponential distribution.  N/A indicates that computation of the inverse of the Hessian was not numerically stable. }
	\label{table:test-exp}
	\vspace*{3mm}
	\begin{tabular}{l l c }
		\hline
		ID &shape parameter  & p-value\\
		\hline
		North Europe &\, 0.04 \; \; N/A & 0.49 \\
		South Europe &-0.04 \; \; (-0.19,  0.10) & 0.19  \\
		Europe &-0.03 \; \; (-0.18,  0.11) &0.52  \\
		North America & -0.03 \; \; (-0.14,  0.09) & 0.31 \\
		Europe\&America & -0.03 \; \; (-0.12, 0.06) & 0.17  \\
		Japan & \, 0.15 \; \; N/A &0.47 \\
		World &-0.03 \; \; (-0.12, 0.06) & 0.21 \\
	\end{tabular}
\end{table}

Estimation using truncated and/or censored observations, such as the truncated observations in the IDL database, is of central interest in survival analysis and in analysis of failure time data, see e.g.  \cite{Andersen-Borgan-Keiding1992} and \cite{Kalbfleish-Prentice2002} for much more information and asymptotic theory. However, the particular truncation setting above does not seem to be considered explicitly in these books. Instead, the formulas are contained in \cite{gampe2010}.

We have not corrected for selection bias in the qq-plots, and in fact, such corrections do not seem to be available in the literature. 

\section{Comments on X. Dong, B. Milholland, J. Vijg,  Nature 538, 257 (2016).}
\label{sec:dong}
\begin{figure}[H]
	\centering
	\includegraphics[width=0.48\textwidth]{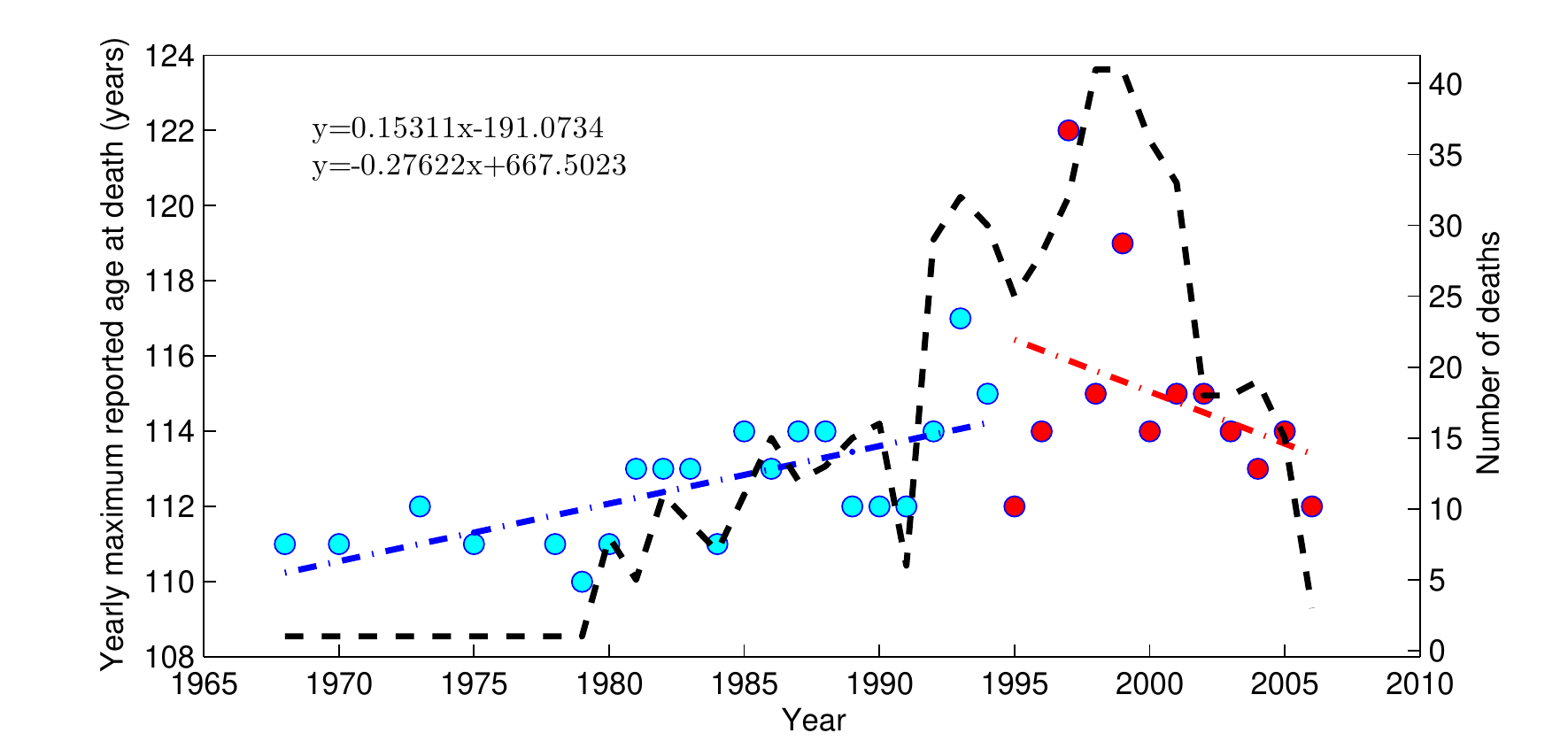}
	\includegraphics[width=0.48\textwidth]{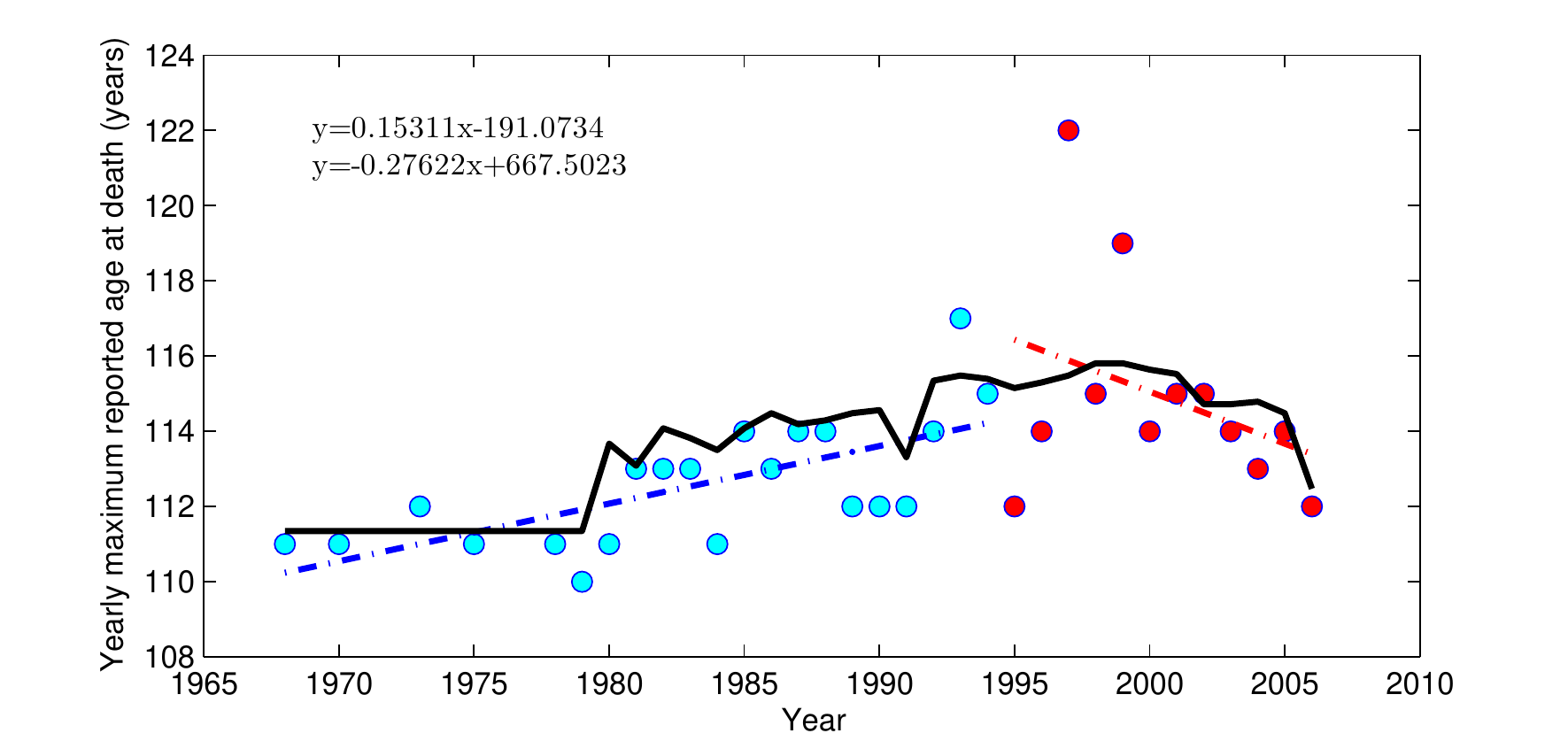}
	\caption{Yearly maximum  reported age at death for supercentenarians. Combined data consisting of England \& Wales 1968-2006, France 1987-2003, USA 1980-2003, and Japan 1996-2005. Dashed - dotted lines:  blue regression lines for  deaths 1968-1994,  red regression line for deaths 1995-2006. {\em Left}: Dashed line is yearly number of deaths, $n_t$. {\em Right}: Solid line is 110 + mean of maximum of $n_t$ exponential variables with $\sigma=1.35$.}
	\label{fig:Dongall}
\end{figure}

Fig.~\ref{fig:Dongall}, left, reproduces Fig.~2.a of the \cite{dongetal2016} Nature Letter, without the truncation of ages to whole years, and with the yearly number of deaths of supercentenarians, $n_t$,  added. The plot is based on  IDL data for  England \& Wales, France, USA, and Japan. The data sets for the different countries cover different time periods, and as a result of this the  yearly number of supercentenarian deaths in the plot varies from 0 to 42. The yearly maximum reported age at death follows the same pattern as the number of supercentenarian deaths, and in particular the decline in the maximum recorded age at death after 1995 follows the decline in yearly number of deaths.  That the yearly maximum reported age at death shows the same pattern as the yearly number of deaths is completely  as expected:  the maximum of many lifespans is stochastically larger than the maximum of few lifespans, and hence the maximum age is likely to be larger for years with many deaths.

\begin{figure}[H]
	\centering
	\includegraphics[width=0.49\textwidth]{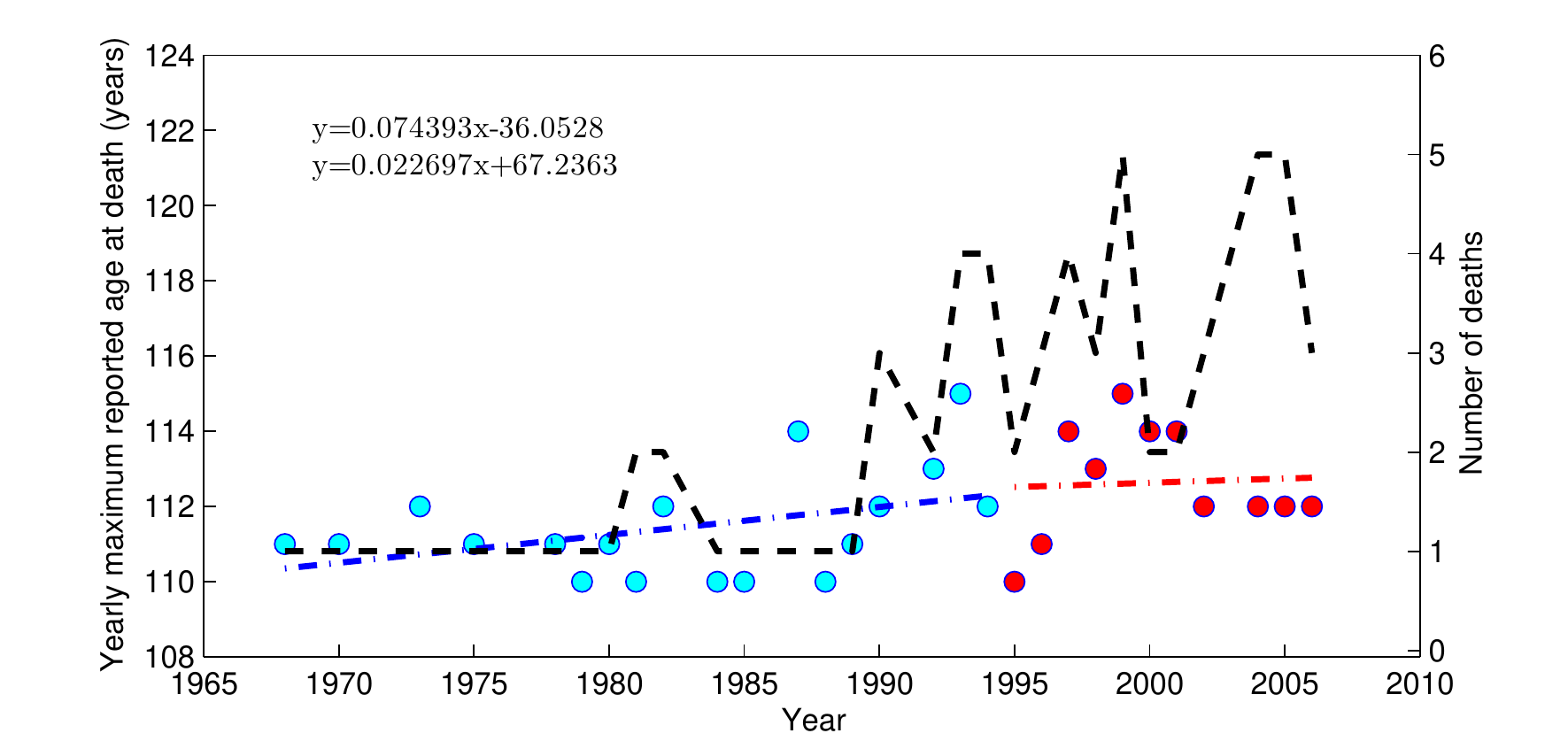}
	\includegraphics[width=0.49\textwidth]{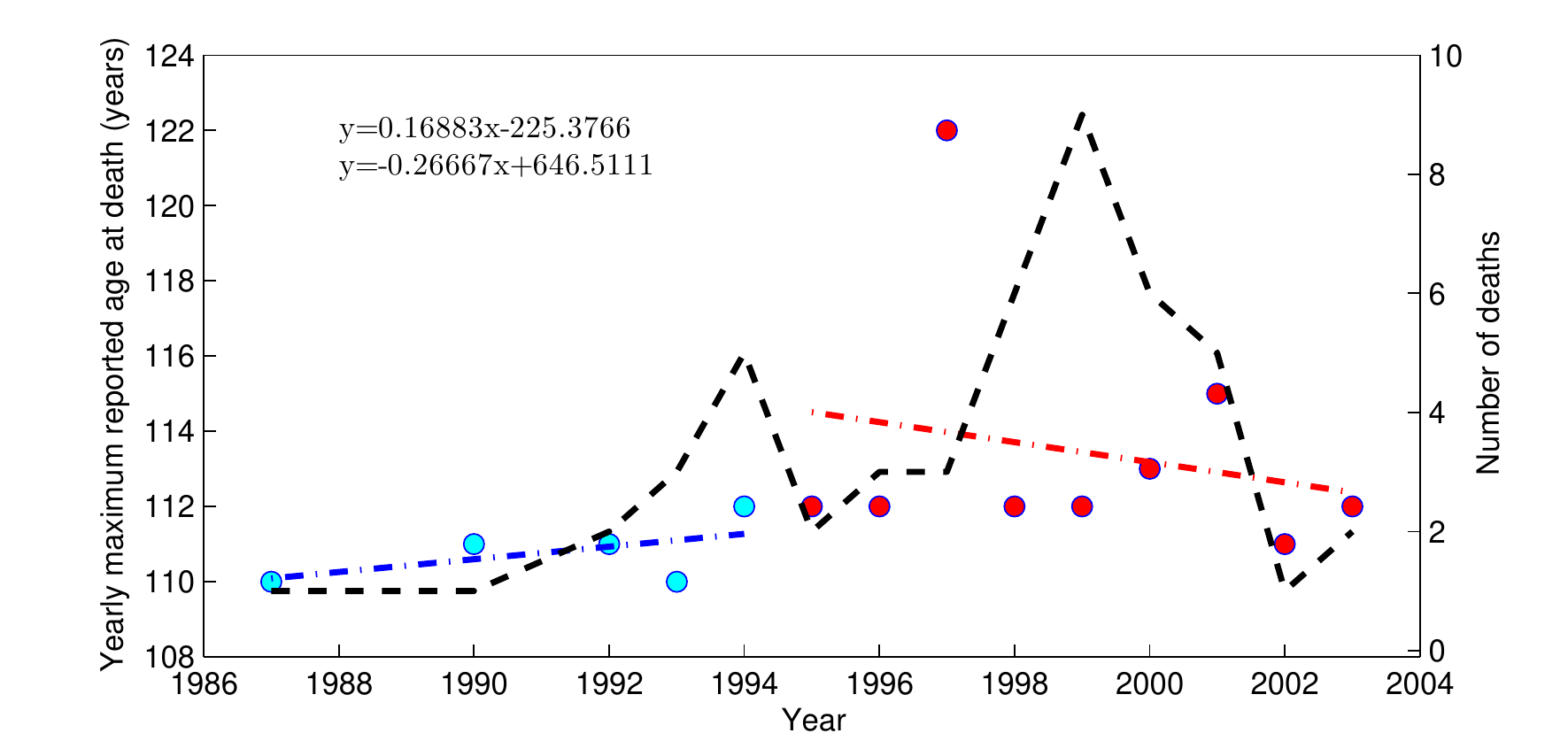}\\
	\includegraphics[width=0.49\textwidth]{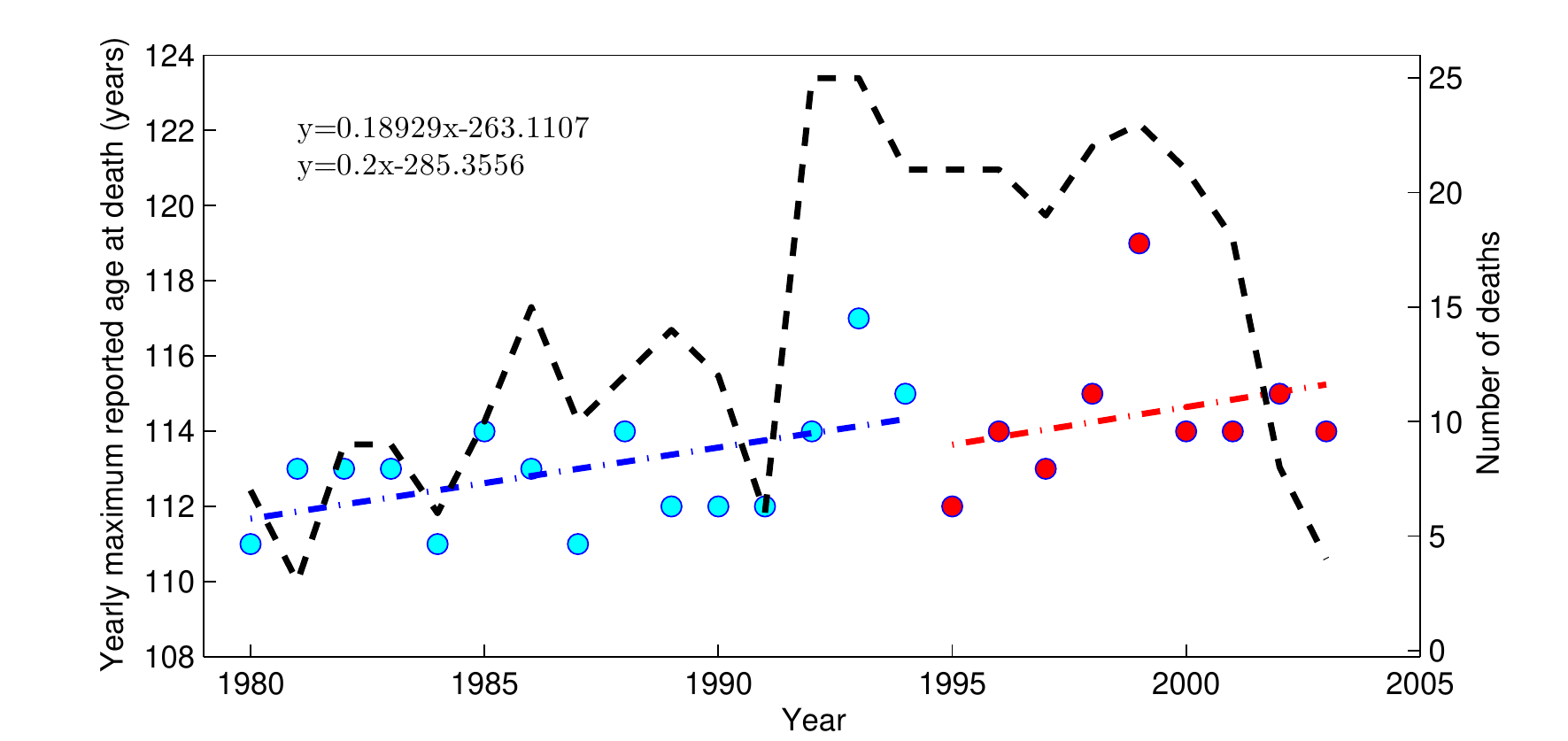}
	\includegraphics[width=0.49\textwidth]{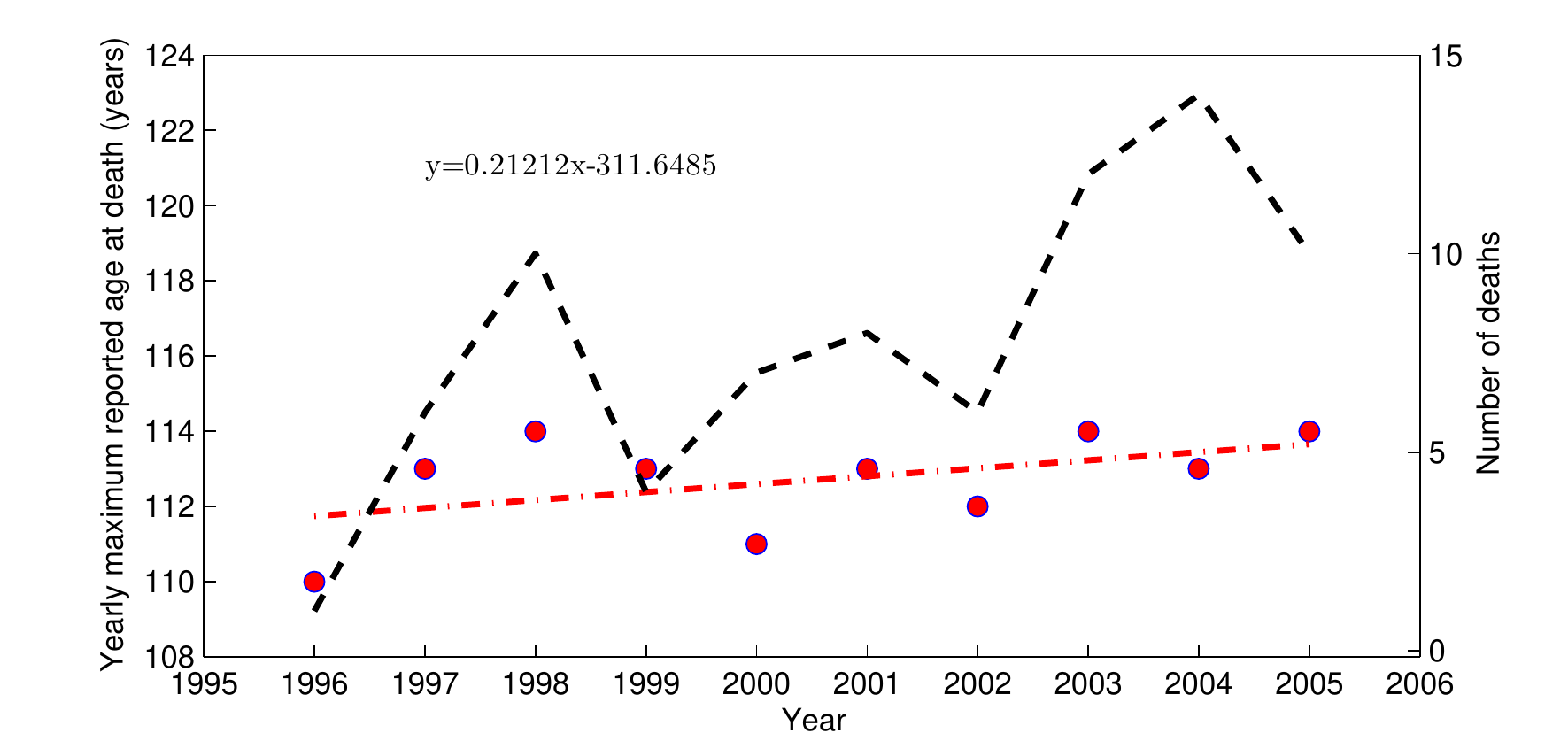}
	\caption{Yearly maximum  reported age at death for supercentenarians.  Dashed - dotted lines:  blue is regression lines for  deaths 1968-1994,  red regression line for deaths 1995-2006. Dashed lines are yearly number of deaths. {\em Top left:} England \&Wales, {\em Top right:} France, {\em Bottom left:} USA, and {\em Bottom right:} Japan.}
	\label{fig:Dong4}
\end{figure}

The right plot in Fig.~\ref{fig:Dongall} instead of $n_t$ adds the mean of the maximum of $n_t$ independent exponential variables, with parameter $1.35$ estimated from these countries.  This does not take truncation into account, but still, again as expected, the mean and the regression lines agree well.   Fig.~\ref{fig:Dong4} shows the same plots as in Fig.~\ref{fig:Dongall}, left,  done separately for individual countries. There is no evidence of a trend break in these plots.
Thus,  in conclusion, the apparent trend break in Fig.~2.a in \cite{dongetal2016} is an artifact caused by inappropriate combination of data from different time periods.

Fig.~2.b of \cite{dongetal2016} adds the 2nd to 5th highest reported age at death to Fig.~2.a. The pattern is the same as in Fig.~2.a, and the explanation is the same:  inappropriate combination of data from different time periods. Further, the statement in \cite{dongetal2016} that ``the annual average age at death for these supercentenarians has not increased since 1968 (Fig.~2.c)'' does not  imply that there is a limit. Again: records increase as more attempts to break them are made, even without a change in the underlying distribution. The conclusion ``our data strongly suggest that the duration of life is limited'' in \cite{dongetal2016} is based on wrong and misleading analysis.

\section{Discussion}
\label{sec:discussion}
The fact that data doesn't  shows any difference in  survival at extreme age between women and men; between persons with different lifestyles or genetic backgrounds; between different time periods; or between different ages, say between 110-year-olds and 115-year-olds is surprising and remarkable and can inform the search for demographic or biological theories of aging. In particular, the fact that lifestyle factors which are important for survival at younger ages ceases to play a role after age 110 is of both biological and popular interest.

Box 1 of \cite{vaupel2010} gives an introduction to the important explanation of mortality at high age as resulting from a mixture of subpopulations with different frailty.
Presumably the composition of subpopulations of humans reaching 110 years of age differs, at least somewhat, between persons who lived earlier and who lived later, or between countries. If the frailty hypothesis is correct, mortality should then be different in at least some of the countries or in one of the time periods. However, we did not find any such differences. This weakens the case for the frailty explanation.
A fundamental hypothesis  on how we  age is  that ``the rate at which the chance of death increases with age for humans is a basic biological constant, very similar and perhaps invariant across individuals and over time'', see \cite{vaupel2010}. For mortality after 110, this would mean that the distribution of excess life length in the future should still be exponential, but with lower yearly probability of death. The methods developed in this paper provide an  efficient way of checking if this indeed is the case: power calculations show that an increase of 5\% or more in yearly survival in a new data set of the same size as the IDL data  would be detectable.

Substantial efforts to find ``a cure for aging''  and in ``engineering better humans'' with the aim of extending the length of human life, perhaps indefinitely, are underway, see e.g. \cite{vijg-campisi:2008,longo-et-al:201,haggstrom2016}. Success of such efforts could happen either in a dramatic and obvious way, or gradually. In the latter case the results of this paper, together with similar analyses  of data to be collected in the future, will aid early confirmation of success of the efforts.

\vspace*{3mm}

\noindent
{\bf Supplementary materials:} LATool - a MATLAB toolbox for life length analysis which makes it possible to obtain more detailed results and do alternative analyses. Available at www.zholud.com.

\vspace*{3mm}
\noindent
{\bf Acknowledgements}

\noindent
We thank Anthony Davison, Jutta Gampe, Olle H\"aggstr\"om, Peter Jagers, Niels Keiding, Thomas Mikosch and Olle Nerman for comments. Research supported by the Knut and Alice Wallenberg Foundation, grant KAW 2012.0067.

\bibliographystyle{chicago}
{\small
	\bibliography{libraryLongevity}
}
\end{document}